\documentclass[twocolumn]{jpsj2} 
\usepackage{bm}

\title{Unconventional Superconductivity Induced by Quantum Critical Fluctuations in Hydrate Cobaltate Na$_{x}$(H$_3$O)$_{z}$CoO$_{2}\cdot$ $y$H$_{2}$O\\
--- Relationship between Magnetic Fluctuations and Superconductivity Revealed by Co Nuclear Quadrupole Resonance --- }

\author{
Y. \textsc{Ihara,}$^{1}$\thanks{E-mail address: ihara@scphys.kyoto-u.ac.jp}
H. \textsc{Takeya,}$^{1}$
K. \textsc{Ishida,}$^{1,2}$\thanks{E-mail address: kishida@scphys.kyoto-u.ac.jp}
H. \textsc{Ikeda,}$^{1}$
C. \textsc{Michioka,}$^{3}$
K. \textsc{Yoshimura,}$^{3}$
K. \textsc{Takada,}$^{4}$
T. \textsc{Sasaki,}$^{4}$
H. \textsc{Sakurai}$^{5}$
and
E. \textsc{Takayama-Muromachi}$^{5}$}

\inst{$^{1}$Department of Physics, Graduate School of Science, Kyoto University, Kyoto 606-8502, Japan \\
$^{2}$ International Innovation Center, Kyoto University, Kyoto 606-8501, Japan \\
$^{3}$Department of Chemistry, Graduate School of Science, Kyoto University, Kyoto 606-8502, Japan \\
$^{4}$Nanoscale Materials Center, National Institute for Materials Science, Tsukuba, Ibaraki, 305-0044, Japan \\
$^{5}$Advanced Nano Materials Laboratory, National Institute for Materials Science, Tsukuba, Ibaraki 305-0044, Japan}

\abst{
Co nuclear-quadrupole-resonance (NQR) measurements were performed on various bilayered hydrate cobaltate Na$_{x}$(H$_3$O)$_{z}$CoO$_{2}\cdot y$H$_{2}$O 
with different values of the superconducting and magnetic-ordering temperatures, 
$T_{\rm c}$ and $T_{\rm M}$, respectively. 
From measurements of the temperature and sample dependence of the NQR frequency, 
it was revealed that the NQR frequency is changed by the change of the electric field gradient (EFG) 
along the $c$ axis $\nu_{\rm zz}$ rather than the asymmetry of EFG within the $ab$-plane. 
In addition, 
it is considered that the change of $\nu_{\rm zz}$ is gaverned mainly by the trigonal distortion of the 
CoO$_{2}$ block layers along the $c$ axis, from the relationships between $\nu_{\rm zz}$ and the various physical parameters. 
We found the tendency that samples with $\nu_{\rm zz}$ larger than 4.2 MHz show magnetic ordering, 
whereas samples with lower $\nu_{\rm zz}$ show superconductivity. 
We measured the nuclear spin-lattice relaxation rate $1/T_1$ in these samples, 
and found that magnetic fluctuations depend on samples.
The higher-$\nu_{\rm zz}$ sample has stronger magnetic fluctuations at $T_{\rm c}$.   
From the relationship between $\nu_{\rm zz}$ and $T_{\rm c}$ or $T_{\rm M}$, 
we suggest that the NQR frequency can be regarded as a tuning parameter to determine the ground state of the system, 
and develop the phase diagram using $\nu_{\rm zz}$. 
This phase diagram shows that the highest-$T_{\rm c}$ sample is located at the point where $T_{\rm M}$ is considered to be zero, 
which suggests that the superconductivity is induced by quantum critical fluctuations. 
We strongly advocate that the hydrate cobaltate superconductor presents an example of 
the magnetic-fluctuation-mediated superconductivity argued in the heavy-fermion compounds. 
The coexistence of superconductivity and magnetism observed in the sample with the highest $\nu_{\rm zz}$ 
is also discussed on the basis of the results of our experiments. 
}

\kword{superconductivity, hydrous sodium cobalt oxide, NQR, spin fluctuations}

\begin{document}
\maketitle

\section{Introduction} \label{Intro}

Since the superconductivity in the hydrate cobaltate Na$_{0.3}$CoO$_{2}\cdot 1.3$H$_{2}$O was discovered\cite{takada}, 
many studies have been conducted from both experimental and theoretical points of view. 
Superconductivity on the two-dimensional triangular lattice CoO$_{2}$ layer is attractive,
because unconventional superconductivity in cuprate and ruthenate is realized on the two-dimensional square lattice.  
In the cobaltate compounds, 
the superconductivity emerges only when water molecules are sufficiently intercalated between the CoO$_{2}$ layers. 
This superconducting structure is called the bilayered hydrate (BLH) structure, in which the double water layers sandwich a Na layer and a water-Na block layer is formed\cite{jorgensen}. 
The BLH compounds are unstable and the sample quality is easily degraded under ambient conditions, since water molecules easily evaporate into air\cite{foo}. 
After the evaporation of some water, the BLH structure changes to a single-layer structure, in which the water-Na block layer is changed to a single water-Na layer\cite{takada2}. 
This is called monolayered hydrate (MLH) structure and it does not show any superconductivity. 
It is considered that water intercalation plays an important role in the occurrence of the superconductivity. 

Several roles of water intercalation have already been pointed out.  
First, two-dimensionality is enhanced by water intercalation. 
Since the $c$-axis lattice parameter in the BLH compound is almost twice as long as that of anhydrate Na$_{x}$CoO$_{2}$, 
the interlayer coupling observed in the anhydrate compounds by neutron scattering measurement\cite{bayrakci,helme} is suppressed by water intercalation\cite{Arita}.
Second, the random occupation of the Na$^{+}$ ions is screened by the double water layers. 
Indeed, a single Co nuclear quadrupole resonance (NQR) peak is observed in the BLH compound, whereas two peaks with a satellite structure are observed in the MLH compound. 
The NQR results indicate that the electric field gradient (EFG) at the Co sites in the CoO$_2$ layer is unique in the BLH compound but the several Co sites with different EFG are present in the MLH compound\cite{ishida1}. 
Third, the compression of the CoO$_{2}$ block layers along the $c$ axis is induced by water intercalation\cite{ihara1}. 
When the $c$-axis lattice parameter is elongated by water intercalation, the coupling between the positive Na$^{+}$ and the negative oxygen O$^{2-}$ ions becomes weaker.
Consequently, the oxygen ions shift toward the positive Co ions, resulting in squeezing of the CoO$_{2}$ block layers. 
The neutron diffraction measurements show that $T_{\rm c}$ increases with decreasing CoO$_{2}$ layer thickness\cite{lynn}. 
Although water content is regarded as an important factor in superconductivity, it is difficult to control water content precisely due to the characteristics of the soft chemical procedure\cite{sakurai1}. 
This is why there exists a large sample dependence of superconductivity in this system. 

Another controversial point is the Fermi-surface (FS) properties in the cobaltate compounds. The LDA band calculation in NaCo$_2$O$_4$ carried out by Singh suggested that the FSs of the compound consist of a large holelike FS composed of an $a_{1g}$-orbital around the $\Gamma$ point and six small hole pockets composed of $e'_{g}$-orbitals around the $K$ points\cite{singh}. 
However, angle-resolved photoemission spectroscopy (ARPES) experiments performed by several groups revealed a large $a_{1g}$-FS and the absence of the small hole pockets. They also found that the $e'_g$ bands and the associated small pockets sink below the Fermi energy\cite{qian, yang, shimojima}, although the bands and the pockets are reported to be raised by water intercalation\cite{shimojima2}. The topology of the FSs in the hydrate and anhydrate cobaltate is still uncertain at the moment, but is important for developing the theoretical scenario for the occurrence of the unconventional superconductivity\cite{mochizuki, yanase, yada}. 
We point out that the investigation of the magnetic fluctuations in anhydrate, MLH and BLH compounds will provide important information about the FS in these compounds, because the magnetic fluctuations are induced by the nesting of the FSs. 

We have performed Co-NQR measurements on various samples with different superconducting and magnetic-ordering temperatures, $T_{\rm c}$ and $T_{\text{M}}$, respectively, in order to investigate the sample dependence of the magnetic fluctuations in the superconducting cobaltate and to identify the key factors for the occurrence of the superconductivity.
We found that the NQR frequency $\nu_{\rm Q}$ and the magnetic fluctuations strongly depend on the sample.
After the investigation of the relationship between the observed $\nu_{\rm Q}$ and various physical parameters, it was found that $\nu_{\rm Q}$ is proportional to the $c$-axis lattice parameter and is related to $T_{\rm c}$. 
In addition, the samples with higher $\nu_{\rm Q}$ show magnetic ordering. 
We consider that $\nu_{\rm Q}$ is one of the important physical quantities for determining the electronic state and develop a phase diagram of the cobaltate superconductor using $\nu_{\rm Q}$.
The phase diagram we show here indicates that the highest-$T_{\rm c}$ sample is situated at the point where the magnetic-ordering temperature becomes zero, which suggests that the superconductivity on the BLH cobaltates is induced by the quantum critical fluctuations. 

\section{Experiment}

We used twelve samples that were synthesized as described in the literature\cite{takada2, sakurai1}. 
The crystalline parameters and Na content $x$ are precisely measured by X-ray diffraction and 
inductively coupled plasma atomic emission spectroscopy (ICP-AES), respectively. 
The Co valence $v$ in several samples is measured by redox titration measurements. 
The superconducting transition temperature $T_{\text{c}}$ is determined from the onset of the superconducting diamagnetism. 
NQR spectra are obtained by recording signal intensity with changing frequency. 
Various physical parameters are listed in Table I.
From the redox-titration measurements, 
it has already been pointed out by several groups that the valance of Co $v$ cannot be interpreted with 
the chemical formula of Na$_{x}$CoO$_{2}\cdot y$H$_{2}$O, 
since the Na content $x$ is not equal to $4-v$\cite{milne,sakurai4}. 
Taking into account the presence of the oxonium ion, H$_{3}$O$^{+}$, 
it has become widely accepted that the chemical formula is expressed as Na$_{x}$(H$_3$O)$_z$CoO$_{2}\cdot y$H$_{2}$O\cite{takada3}. 
Here, the oxonium content is determined to be $z = 4-x-v$.
Nuclear spin-lattice relaxation rate $1/T_1$ was measured at the peak maximum of the NQR signal 
arising from $\pm 5/2 \leftrightarrow \pm 7/2$ transitions in each sample. 
The recovery of the nuclear magnetization after saturation pulses can be fitted by the theoretical curves 
in the whole temperature range.
To obtain reliable data and maintain the sample quality during the measurements, 
all samples are preserved in freezer or in liquid N$_{2}$, 
and thus the superconducting and magnetic transition temperatures and NQR results in each sample are reproducible.

\begin{table}[htp]
\caption{Na content, $c$-axis length, transition temperature, Co valence and NQR frequency arising from the transition between $\pm5/2 \leftrightarrow \pm7/2$ .} 
\label{table:NCOsample}
\begin{tabular}{rcccccc}\hline
No.& Na content & $c$ (\AA) & $T_{\rm c}$ (K) & Co$^{+v}$ & $\nu_{3}$ (MHz) & ref. \\ \hline
 1 & 0.348 & 19.684 & 4.7  & ---  & 12.30 & \cite{ishida1}\\
 2 & 0.339 & 19.720 & 4.6  & ---  & 12.30 & \cite{ihara4}\\
 3 & 0.348 & 19.569 & 2.8  & ---  & 12.07 & \cite{ihara1}\\
 4 & 0.35  & 19.603 & 4.6  & ---  & 12.32 & \cite{ihara1}\\
 5 & 0.331 & 19.751 & $<1.5$  & 3.42 & 12.54 & \cite{ihara1}\\
 6 & 0.351 & 19.714 & 4.6  & 3.37 & 12.45 & \cite{ihara3,ihara5}\\
 7 & 0.350 & 19.739 & 4.4  & 3.41 & 12.50 & \\
 8 & 0.322 & 19.820 & 3.6  & 3.40 & 12.69 & \\
 9 &  ---  &  ---   & 3.6  & ---  & 12.22 & \\
10 & 0.380 & 19.588 & 2.6  & 3.49 & 12.10 & \\
11 & 0.368 & 19.614 & 3.5  & 3.47 & 12.18 & \\
12 & 0.346 & 19.691 & 4.8  & 3.48 & 12.39 & \\ \hline
\end{tabular}
\end{table}

\section{Experimental Results}

\subsection{NQR spectrum}

\begin{figure}
\begin{center}
\includegraphics[width=8cm]{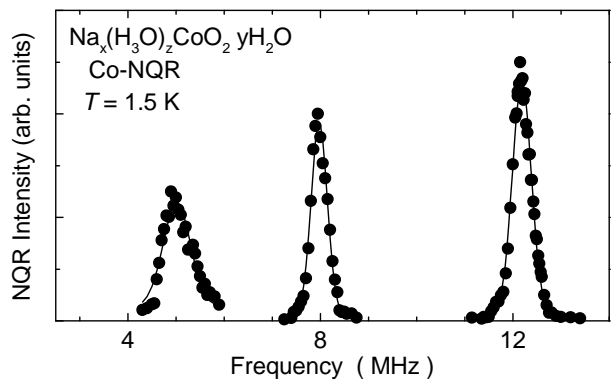}
\end{center}
\caption{
Co-NQR spectra at 1.5 K in sample No. 11. These peaks originate from three transitions.
From these peak frequencies, $\nu_{\rm Q}$ and $\eta$ are evaluated to be 4.08 MHz and 0.20, respectively.}
\label{fig1}
\end{figure}
The nuclear Hamiltonian with the nuclear electric quadrupole interaction is described as
\[
{\cal H}_{\rm Q} = \frac{\nu_{\rm zz}}{6}\left[3I_z^2-{\bm I}^2+\frac{\eta(I_+^2+I_-^2)}{2}\right], 
\]
 where quadrupole frequency $\nu_{\rm zz}$ is defined as $3e^2qQ/2I(2I-1)$ with the electric field gradient (EFG) along the $z$ axis $eq = V_{\rm zz}$ and the nuclear quadrupole moment $Q$. The asymmetry parameter of EFG $\eta$ is expressed as $(V_{\rm xx}-V_{\rm yy})/V_{\rm zz}$ with $V_{\alpha \alpha}$, which is EFG along the $\alpha$ direction ($\alpha=x,y$ and $z$).
The nuclear spin of Co is $I = 7/2$, and three NQR peaks are observed from $\pm 1/2 \leftrightarrow \pm 3/2$, $\pm 3/2 \leftrightarrow \pm 5/2$, and $\pm 5/2 \leftrightarrow \pm 7/2$ transitions. The typical NQR spectrum in a superconducting cobaltate is shown in Fig. 1. 
Because of the existence of an asymmetry of EFG in the $ab$ plane, the separation between two peaks is not the same. 
Within the second-order perturbation against the asymmetry parameter in the $ab$ plane $\eta$, resonant frequencies are described as
\begin{eqnarray}
   \nu_{1}&=&\nu_{\rm zz}\left(1 + \frac{109}{30}\eta^{2}\right)\nonumber
~~(\mbox{for} \pm 1/2 \leftrightarrow \pm 3/2),\\
   \nu_{2}&=&2\nu_{\rm zz}\left(1 - \frac{17}{30}\eta^{2}\right)\nonumber
~~(\mbox{for} \pm 3/2 \leftrightarrow \pm 5/2),\\
   \nu_{3}&=&3\nu_{\rm zz}\left(1 - \frac{1}{10}\eta^{2}\right) \nonumber
~~(\mbox{for} \pm 5/2 \leftrightarrow \pm 7/2).
\end{eqnarray}
Both $\nu_{\rm zz}$ and $\eta$ are obtained by solving the above equations. 
$\nu_{\rm zz}$ is determined by several factors, such as the Co valence, the hole density of the Co $3d$ orbitals and the crystal structure, particularly the distance between O$^{2-}$ and Co$^{v+}$ ions. 

\begin{figure}
\begin{center}
\includegraphics[width=8cm]{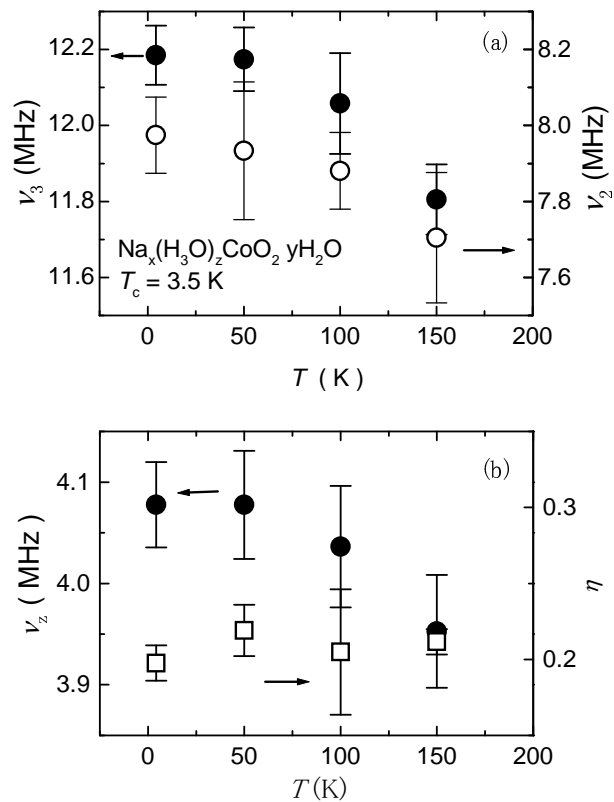}
\end{center}
\caption{(a) Temperature dependence of $\nu_{3}$ and $\nu_{2}$ measured on sample No. 11. (b) Temperature dependences of $\nu_{\rm zz}$ and $\eta$, which were evaluated from $\nu_{2}$ and $\nu_{3}$ in (a).}
\label{fig2}
\end{figure}%
In order to investigate the temperature dependence of $\nu_{\rm zz}$ and $\eta$, 
both $\nu_{2}$ and $\nu_{3}$ are measured up to 150 K in sample No. 11. 
As seen in Fig. \ref{fig2} (a), both $\nu_{2}$ and $\nu_{3}$ are nearly constant below 50 K, 
and shift toward a lower frequency with increasing temperature.
Temperature dependences of $\nu_{\rm zz}$ and $\eta$, 
which are derived from the temperature variations of $\nu_{2}$ and $\nu_{3}$, are shown in Fig.~\ref{fig2}~(b). 
We found that the temperature dependence of $\nu_{\rm zz}$ is similar to that of the $c$-axis lattice parameter 
measured from the neutron diffraction\cite{lynn}, and suggest that the temperature dependence of $\nu_{\rm zz}$ 
is mainly determined by that of the $c$-axis parameter \cite{lynn}. 
In contrast to the considerably strong temperature dependence of $\nu_{\rm zz}$, 
$\eta$ is independennt of temperature and is estimated to be $\sim$ 0.2. 
Because $\eta$ originates from the anisotropy of EFG along $x$ and $y$ directions, 
it should be noted that $\eta$ is zero in a perfect triangular lattice. 
The finite value of $\eta$ suggests some distortions are present in the $x$-$y$ direction of the CoO$_2$ layer.
The possible origin of $\eta$ will be discussed in \S \ref{EFG}. 

\begin{figure}
\begin{center}
\includegraphics[width=8cm]{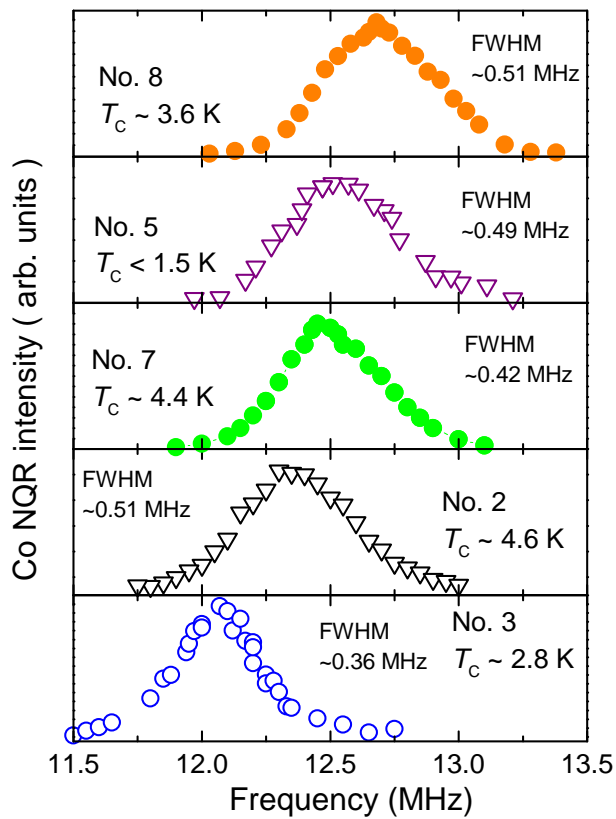}
\end{center}
\caption{Co-NQR spectra arising from the $\pm5/2 \leftrightarrow \pm7/2$ transitions in various samples measured at 4.2 K. In magnetically ordered samples (No. 5 and No. 8), the spectra were measured at 8 K above ordering temperatures.
$T_{\rm c}$ and the full width at half maximum in each sample are shown.
}
\label{NQR}
\end{figure}%

Figure \ref{NQR} shows the NQR spectra arising from the $\pm 5/2 \leftrightarrow \pm 7/2$ transition in various samples with different contents of H$_{2}$O, Na$^{+}$ and H$_{3}$O$^{+}$. 
The values of $\nu_{\rm zz}$ and $\eta \equiv (\nu_{\rm xx}-\nu_{\rm yy})/\nu_{\rm zz}$ are evaluated in these samples. 
The values of $\nu_{\rm zz}$ and $\nu_{\rm xx}-\nu_{\rm yy}(=\eta \times \nu_{zz})$ are plotted against $\nu_{3}$ in Fig. 4 (a).  
The value of $\nu_{\rm xx}-\nu_{\rm yy}$ is nearly constant, 
in contrast to the linear relationship between $\nu_{\rm zz}$ and $\nu_{3}$. 
This indicates that the sample dependence of NQR frequency $\nu_{3}$ in our various samples originates mainly from $\nu_{\rm zz}$. 
Recently, Zheng {\it et al.} reported that $\eta$ exhibits strong sample dependence in samples with different Na contents ($0.26 < x < 0.34$)\cite{zheng1}. The variation of the Na content in our samples is smaller ($0.322 < x < 0.38$ ), although the $T_{\rm c}$ variation is larger than that in their sample. It seems that $\nu_{xx}-\nu_{yy}$ is sensitive to the Na content\cite{zheng1}.     
The $\nu_{\rm zz}$ is plotted against various physical parameters such as Na content $x$ and Co valence $v$ (Fig. 4 (b)), and the oxonium ion content $z$ and the $c$-axis parameter (Fig. 4 (c)), where the oxonium ion content $z$ is estimated from $z=4-x-v$. Detailed discussion on the electric field gradient will be given in \S \ref{EFG}.

\begin{figure}
\begin{center}
\includegraphics[width=8cm]{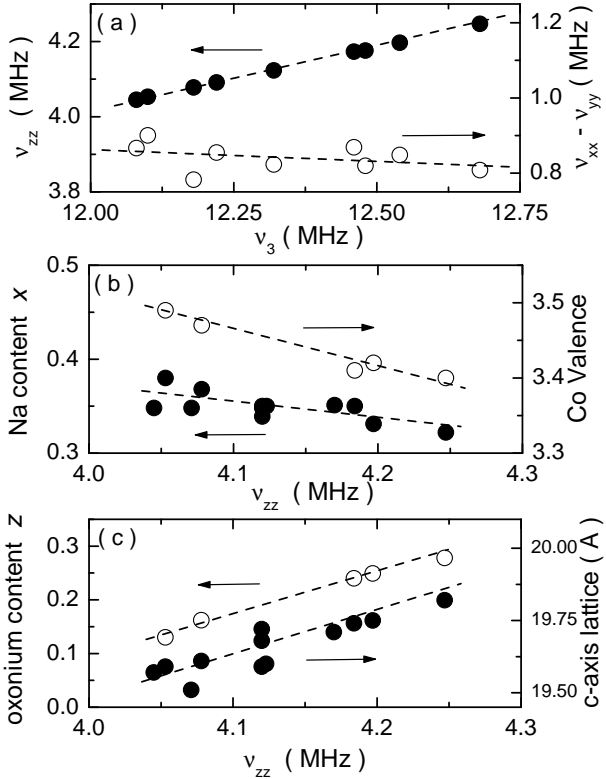}
\end{center}
\caption{(a) $\nu_{\rm zz}$ and $\nu_{\rm xx}-\nu_{\rm yy}$ are plotted against $\nu_{3}$. (b) and (c) Various physical parameters (Na content: $x$, Co valence: $v$, oxonium content: $z$, and c-axis lattice parameter) are plotted against $\nu_{\rm zz}$.}
\label{fig3}
\end{figure}%

\subsection{Nuclear spin-lattice relaxation rate}

\begin{figure}
\begin{center}
\includegraphics[width=8cm]{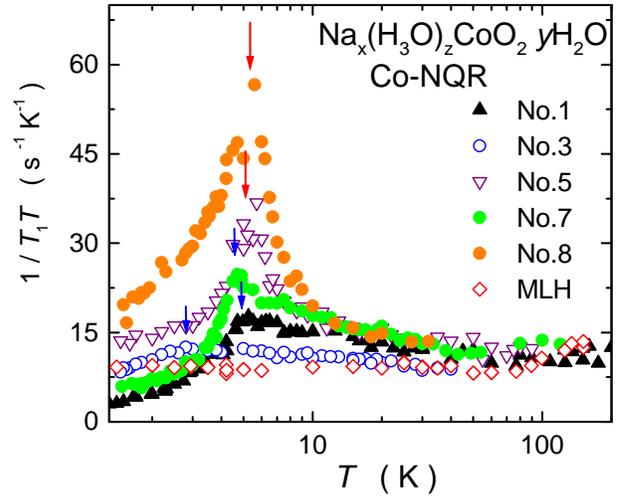}
\end{center}
\caption{Temperature dependence of $1/T_1T$ in various samples. The blue (red) arrows indicate $T_{\rm c}$ ($T_{\rm M}$). The values of $1/T_1T$ at $T_{\rm c}$ (or just above $T_M$) increase as the NQR frequency becomes higher. }
\label{invT1T}
\end{figure}%

Figure \ref{invT1T} shows the temperature dependence of $1/T_1T$ in various BLH samples below 200 K, together with that in the MLH sample which does not show superconductivity down to 1.5 K. 
The values of $1/T_1T$ in all samples show similar temperature dependences down to 70 K. 
Below 70 K, $1/T_1T$ shows a sample dependence. The non-superconducting MLH sample shows the Korringa behavior down to 1.5 K, which is indicative of the absence of temperature-dependent magnetic fluctuations, 
whereas the values of $1/T_1T$ in superconducting samples (No. 1, 3 and 7) increase with decreasing temperature below 70 K down to $T_{\rm c}$. Blue arrows in Fig.~\ref{invT1T} indicate $T_{\rm c}$ of these samples. In magnetic samples (No. 5 and 8), $1/T_1T$ shows a prominent peak at $T_{\rm M}$, shown by red arrows, below which an internal field appears at the Co nuclear site. The ordered moment in sample No. 5 was estimated to be on the order of 0.01 $\mu_{\rm B}$ using the observed internal field strength and the hyperfine coupling constant at the Co nuclear site\cite{ihara1}. 
It is noteworthy that the temperature dependence of $1/T_1T$ in the high-$T_{\rm c}$ sample (No. 1) is the same as that in the magnetic samples in the temperature range of 15 - 100 K. This suggests that the high-$T_{\rm c}$ samples have similar magnetic fluctuations as those in the samples with magnetic ordering.

In addition, the low-$T_{\rm c}$ sample (No. 3) shows a weak enhancement of $1/T_1T$ down to $T_{\rm c}$, followed by the moderate decrease of $1/T_1T$ below $T_{\rm c}$, although the full width at half maximum (FWHM) of the NQR peak is the narrowest within all the samples. These results suggest that the origin of lower $T_{\rm c}$ is not the suppression of the superconductivity by the inhomogeneity and/or randomness of the sample, but the weakness of the pairing interaction. If $T_{\rm c}$ were suppressed by the inhomogeneity, the FWHM of lower-$T_{\rm c}$ sample would be broader than that of the higher-$T_{\rm c}$ sample. Taking these experimental results into account, we strongly advocate that the magnetic fluctuations, which are close to magnetic instability, play an important role in the superconductivity in BLH cobaltate.
We also point out, on the basis of the comparison between Fig. \ref{NQR} and Fig. \ref{invT1T}, that $1/T_1T$ below 70 K is enhanced more significantly as the NQR frequency becomes higher, and that there is a relationship between low-temperature magnetic fluctuations and the NQR frequency.  
The detailed characteristics of the magnetic fluctuations and the possible interpretation for the relationship are discussed in \S \ref{relaxationrate} and \S \ref{relationship}.
 
We observed the coexistence of magnetic ordering and superconductivity in sample No. 8, which has the highest NQR frequency ($\nu_3$ = 12.69 MHz). The magnetic ordering is identified by the prominent peak of $1/T_1T$, and the occurrence of superconductivity is confirmed by the superconducting diamagnetic signal. The coexistence observed in sample No. 8 will also be discussed in \S \ref{coexistence}.

\section{Discussion}

\subsection{Analyses of EFG at Co site}\label{EFG}

As shown in Fig. \ref{NQR}, EFG at the Co site depends on the sample. Here, we discuss the origin of the sample dependence of EFG.   
In general, there are two sources of EFG. 
One is the point charges on the lattice ions surrounding the Co site, 
$V_{\text{lat}}$, and the other is the intrasite effect $V_{\text{int}}$ produced by the on-site Co-$3d$ electrons.
As shown in Fig. 4 (c), 
it is obvious that there exits some relationship between $\nu_{\rm zz}$ and the $c$-axis lattice parameter. 
It has been reported that the $c$-axis lattice parameter is increased when the water molecules are inserted 
and/or the oxonium ions with the larger ionic radius are replaced by the Na ions\cite{milne, sakurai2}. 
In both cases, 
the block layer consisting of Na/H$_3$O and H$_2$O elongates along the $c$ axis and pushes the CoO$_2$ layer, 
so that the CoO$_2$ block layers are compressed in the samples with the longer $c$ axis. 
It has been pointed out that the thickness of the CoO$_2$ block layer in the BLH compounds with large $y$ values is 
thinner than that of the MLH compounds\cite{sakurai3}.      
According to the point charge calculations, $V_{\text{lat}}$ increased as the CoO$_2$ block was compressed along the $c$ axis. 
The relationship between $\nu_{\rm zz}$ and the $c$-axis lattice parameter is explained consistently in this manner. 

Next, we discuss the contributions from the on-site $3d$ holes. For the cuprate superconductors, it was pointed out that the NQR frequencies at the Cu and O sites are determined by the on-site holes\cite{Hanzawa}, and Zheng {\it et al.} estimated the local hole content from the analyses of $\nu_{\rm Q}$ in those sites\cite{zheng2}. We employ their analyses to cobaltate and evaluate the hole number at Co-$3d$ in the superconducting cobaltate.

When we take $a_{1g}$ and doublet $e'_g$ orbitals in the trigonal structure,   
we describe the wave function of Co-$3d$ orbitals, which consist of the linear combination of those orbitals, as
\[
\psi_{3d} = \sqrt{1-w^2}|a_{1g}> + \frac{w}{\sqrt{2}}|e'_{g+}> + \frac{w}{\sqrt{2}}|e'_{g-}>.
\]
Here, $w^2$ is the occupation of the $e'_g$ orbitals. 
The EFG $V_{\alpha,\beta}$ arising from the on-site holes is calculated as
\[ 
V_{\alpha,\beta} = e\int\psi_{3d}^*~\frac{\partial^2}{\partial\alpha\partial\beta}\left(\frac{-1}{r}\right)~\psi_{3d}~d\tau \hspace{0.3cm} (\alpha, \beta = x,y ~\mbox{and}~z). 
\]
Therefore, each component of $\nu_{\alpha\alpha}$ at the Co site is expressed as 
\[
\nu_{\alpha,\alpha} = n_{3d} \nu_{3d,0} g(w), 
\]
where $n_{3d} = n_{a_{1g}} + n_{e'_{g}}$ is the hole number in the Co-$3d$ $t_{2g}$ orbitals and $g(w)$ denotes the solutions of the secular equation with respect to $V_{\alpha,\beta}$.
The largest value of the solution of $g(w)$ gives the value of $\nu_{\rm zz}$, and the smallest one gives $\nu_{\rm yy}$. 
When there is one hole in the $a_{1g}$ orbital ($w=0$), $\nu_{3d,0}$ is expressed as
\[
\nu_{3d,0}=\frac{1}{14}\frac{e^2~^{59}Q}{h}\frac{1}{4\pi\epsilon}\frac{4}{7}\langle r^{-3}\rangle_{3d},
\]
and it is estimated that $\nu_{3d,0}$ = 20.4 MHz and $\eta$ is zero. 
Here, the average of $r^{-3}$ for the $3d$-orbitals is taken to be $\langle r^{-3}\rangle_{3d}$= 6.7 a.u. and the reduction factor $\xi=0.8$ is adopted\cite{Ray,Mukhamedshin}.

For simplicity, if we neglect $V_{\text{lat}}$ related to the ions surrounding the Co, and assume that the observed $\nu_{\rm Q}$ is determined solely by the on-site Co-$3d$ hole, $w^2 = 0.25$ and $n_{3d} = 0.31$ are derived from the experimental values of $\eta =$ 0.2 and $\nu_{\rm zz} = 4.1$ MHz.
These values imply that the ratio between $n_{a_{1g}}$ and $n_{e'_g}$ is approximately 3 : 1, and the valence of Co is +3.31, which is close to the directly measured $v$ ($v \sim$ 3.45). 
We point out that the non-negligible $\eta$ is due not solely to the Na-ordering effect or the distortion of the $x$-$y$ plane, but also to the presence of holes in the $e'_g$ orbitals, to some degree, because $\eta$ is zero when no hole exists in the $e'_g$ orbitals.  
In addition, if we proceed with this analysis at the O site in the CoO$_2$ layer, the hole content at the O$_{2p}$ $\sigma$ orbitals, which are coupled with the Co-$3d$ $e'_g$ orbitals, is evaluated to be $n_{2p,\sigma} \sim $ 0.05. This is estimated from the O-NMR value ($\nu_{\rm zz} \sim 0.168$ MHz) and $\nu_{2p,0} = - 3.65$ MHz with $\langle r^{-3}\rangle_{2p}$ = 4.97 a.u., using the relation $n_{2p,\sigma} = (1+\eta/3)\nu_{\rm zz}/\nu_{2p,0}$\cite{zheng2}. 

Although the estimated values of the hole content seem to be reasonable, the dependence of $\nu_{\rm zz}$ at the Co site against the directly measured $v$ is inconsistent with the results expected from the analyses based on the on-site holes in the CoO$_2$ layer, because experimental results show that  $\nu_{\rm zz}$ at the Co site increases with decreasing $v$. If the dependence of $\nu_{\rm zz}$ were determined by the change in the hole content, $\nu_{\rm zz}$ should decrease with decreasing $v$. In order to understand the change in $\nu_{\rm zz}$ at the Co site with respect to $v$ completely, the change in $\nu_{\rm zz}$ at the O site should be investigated, since the doped hole might be introduced predominantly into the O-$2p$ orbitals. 
Alternatively, it is possible that the change in $\nu_{\rm zz}$ at the Co site can be qualitatively understood as the change in $V_{\rm lat}$, as discussed above. It seems that the value of $\nu_{\rm zz}$ is predominantly determined by the on-site hole contribution, but the change in $\nu_{\rm zz}$ is predominated by the lattice contribution. We suggest that the observed $\nu_{\rm zz}$ is determined not solely by $V_{\rm int}$, but by both $V_{\rm lat}$ and $V_{\rm int}$. 

\subsection{Spin dynamics in cobaltate compounds}\label{relaxationrate}

\begin{figure}
\begin{center}
\includegraphics[width=8cm]{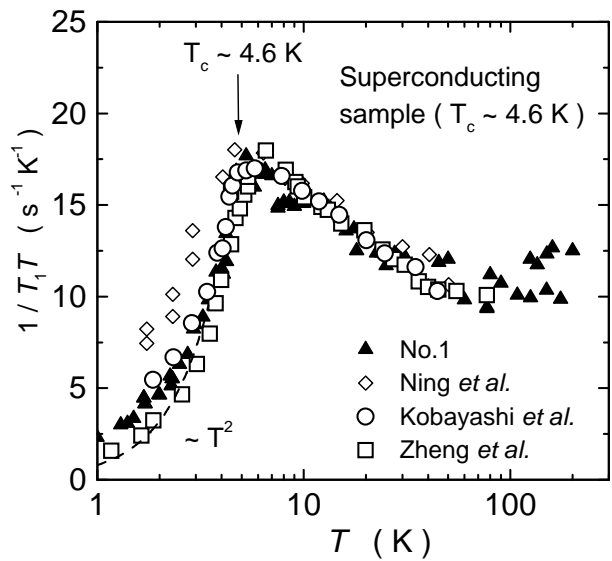}
\end{center}
\caption{Temperature dependence of $1/T_1T$ in superconducting samples with $T_{\rm c} \sim$ 4.6 K reported so far. Temperature dependence of $1/T_1T$ in our sample No.~1 ($T_{\rm c} = 4.7$ K) is compared with that in the superconducting samples with $T_{\rm c} \sim 4.6$ K reported by other groups.\cite{ning1,kobayashi1,fujimoto} }
\label{VariousHTcSample}
\end{figure}

As shown in Fig. \ref{invT1T}, 
$1/T_1T$ in the BLH samples exhibit a strong sample dependence below 70 K. 
In this section, 
we discuss the characteristics of the magnetic fluctuations in the BLH compounds. 
First, the behavior of $1/T_1T$ in sample No. 1 ($T_{\rm c}$ = 4.7 K) is compared with 
that in the samples with $T_{\rm c} \sim 4.6$ K reported by other groups\cite{ning1, kobayashi1, fujimoto}, 
as shown in Fig. \ref{VariousHTcSample}. 
It is obvious that $1/T_1T$ in all samples with $T_{\rm c} \sim 4.6$ K reported so far shows identical behavior 
in the normal state, 
indicating that the temperature dependence of $1/T_1T$ in the normal state is intrinsic in the superconducting samples. 
The possibility that the enhancement of $1/T_1T$ in a low-temperature region between 70 K and 
$T_{\rm c}$ is due to magnetic fluctuations caused by impurity moments is ruled out, 
since the content of impurity moments would differ among samples. In the superconducting state, 
it is clear that $1/T_1$ does not have a coherence peak just below $T_{\rm c}$, 
and shows the $T^3$ dependence ($1/T_{1}T \sim T^2$) down to 1 K. 
This strongly suggests that the superconducting gap has line nodes, 
resulting in the unconventional superconductivity realized in BLH cobaltate\cite{fujimoto,ishida1}.            

\begin{figure}
\begin{center}
\includegraphics[width=8cm]{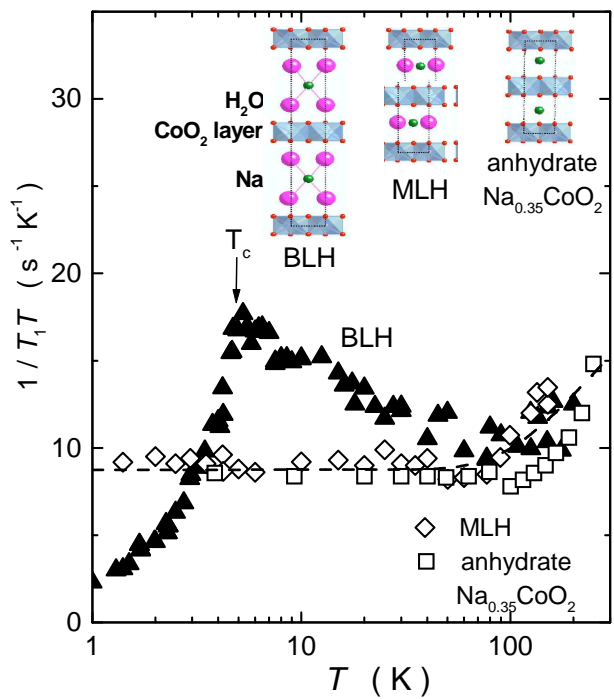}
\end{center}
\caption{Temperature dependence of $1/T_1T$ in superconducting bilayered hydrate (BLH) sample (No. 1) and  monolayered hydrate (MLH) and anhydrate cobaltates Na$_{0.35}$CoO$_2$. The experimental data of the anhydrate cobaltate is taken from Ning and Imai\cite{ning1}. The dashed line is the fitted curve of $1/T_{1}T$ in the MLH and anhydrate compounds (see text).}
\label{BLHMLHT1}
\end{figure}

Next, we compare $1/T_1T$ in the superconducting BLH sample (No. 1) with $1/T_1T$ in the non-superconducting MLH sample\cite{ishida1} and anhydrate Na$_{0.35}$CoO$_2$\cite{ning1}. Figure \ref{BLHMLHT1} shows $1/T_1T$ in these samples. $1/T_1T$ in the MLH sample is almost the same as that in the anhydrate sample, both of which show the Korringa behavior in the low-temperature region. In contrast, $1/T_1T$ in the superconducting BLH compound is enhanced below 70 K down to $T_{\rm c}$, although $1/T_1T$ above 100 K is nearly the same as that in the other two samples. The results of the comparison of the BLH compound with the MLH and anhydrate compounds suggest that the low-temperature enhancement of $1/T_1T$ plays an important role in the occurrence of the superconductivity. In the temperature region greater than 100 K, $1/T_1T$ in all samples increases as temperature increases, which is reminiscent of the spin-gap behavior in cuprate superconductors. It is considered that the spin-gap behavior of $1/T_1T$ in the high-temperature region is related to the pseudogap revealed by the photoemission experiment\cite{shimojima}. A similar spin-gap behavior was reported in anhydrate Na$_x$CoO$_2$ with $x \leq 0.7$\cite{ning2,Yokoi}. The spin-gap behavior followed by the $T_1T = $ const. relation in the MLH and anhydrate samples in Fig.~\ref{BLHMLHT1} is consistently reproduced by 
\[
\left(\frac{1}{T_1T}\right)_{\rm MLH} = 8.75 + 15\exp\left(-\frac{\Delta}{T}\right)~~ \mbox{(sec}^{-1} \mbox{K}^{-1} \mbox{)}
\]
with $\Delta$ = 250 K.
The value of $\Delta$ is in good agreement with the pseudogap of an energy scale of $\sim 20$ meV revealed by the photoemission measurement\cite{shimojima}.
It seems that the spin-gap behavior is a common feature in cobaltate compounds and is unrelated to superconductivity. 
Rather, we suggest that the low-temperature enhancement of $1/T_1T$ is strongly related to superconductivity. The properties of the low-temperature spin dynamics are discussed quantitatively.     

\begin{figure}
\begin{center}
\includegraphics[width=8cm]{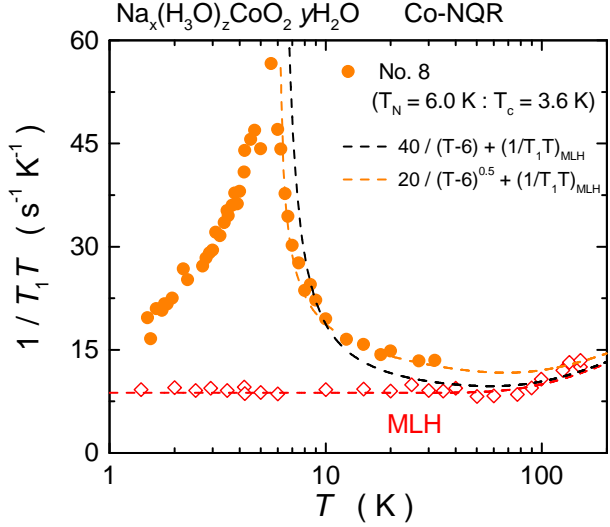}
\end{center}
\caption{Temperature dependence of $1/T_1T$ in sample No. 8, which shows the magnetic order at 6 K.
$1/T_{1}T$ in the MLH compound is also shown. 
The fitted function for the MLH-compound data is shown by the red dashed line. 
The orange and black dashed lines are the fitted functions for data of sample No. 8 above $T_{\rm M}$.}
\label{FittingNo8}
\end{figure}

In order to characterize magnetic fluctuations in the BLH compounds, we first analyze the temperature dependence of $1/T_1T$ in the ordered samples. Figure \ref{FittingNo8} shows the temperature dependence of $1/T_1T$ in sample No. 8, which has the highest NQR frequency and shows magnetic ordering at $T_{\rm M} \sim$ 6 K. In the figure, $1/T_1T$ in the MLH compound is also shown. A prominent divergence of $1/T_1T$ is observed at $T_{\rm M} \sim 6$ K in sample No. 8. We fitted the temperature dependence of $1/T_1T$ above $T_{\rm M}$ with various functions, and found that a fairly good fit can be obtained by a function consisting of two contributions expressed as
\[
\left(\frac{1}{T_1T}\right)_{\rm No.8} = \left(\frac{1}{T_1T}\right)_{\rm MLH} + \frac{20}{\sqrt{T-6}}~~ \mbox{(sec}^{-1} \mbox{K}^{-1} \mbox{)}.
\]
The first term on the right-hand side is the contribution from the spin dynamics observed in the MLH and anhydrate compounds, and the second term is the magnetic contribution, which gives rise to the magnetic ordering. It should be noted that the relation of $1/T_1T \propto 1/\sqrt{T-T_{\rm M}}$ is the temperature dependence anticipated in the three-dimensional (3-D) itinerant antiferromagnet in the framework of the self-consistently renormalized (SCR) theory\cite{Moriya}. It seems that the relation of $1/T_1T \propto 1/(T-T_{\rm M})$ for the 2-D antiferromagnet would be more appropriate if the crystal structure of the BLH compound were taken into account, but the experimental data cannot be reproduced by the function of $1/T_1T \propto 1/(T-T_{\rm M})$, as shown in Fig. \ref{FittingNo8}.
Judging from the temperature dependence of $1/T_1T$, the ordered sample might have the 3-D magnetic correlations just above $T_{\rm M}$ owing to the magnetic ordering, even if the dominant fluctuations are of a 2-D nature.
Nevertheless, we adopt the fitting function of 
\[
\left(\frac{1}{T_1T}\right) = \left(\frac{1}{T_1T}\right)_{\rm MLH} + \frac{a}{\sqrt{T-\theta}}
\]
to understand the systematic change of the magnetic fluctuations.

\begin{figure}
\begin{center}
\includegraphics[width=8cm]{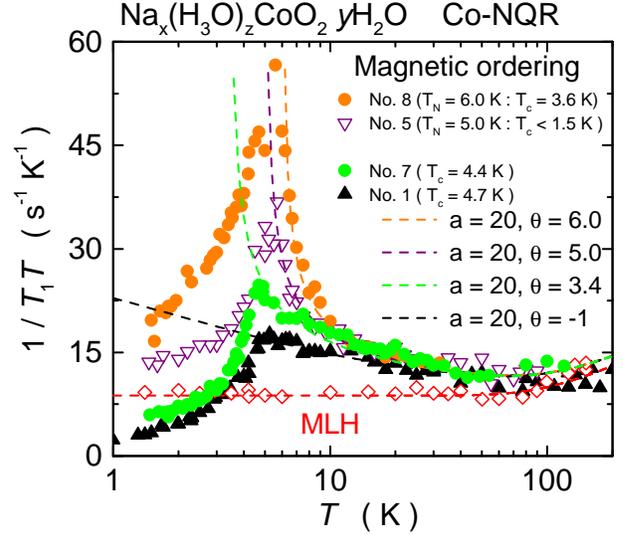}
\end{center}
\caption{Temperature dependence of $1/T_1T$ in the magnetically ordered samples (No. 5 and 8), together with those in the superconducting samples (No. 1 and 7) and the MLH sample. The dashed lines are the fitted results, and the fitting parameters in each sample are shown in the figure (see text).}
\label{MagneticOrderT1}
\end{figure}

Figure \ref{MagneticOrderT1} shows the temperature dependence of $1/T_1T$ in samples No.~5, 7, and 8, together with that in No. 1 and the MLH compounds. Samples No.~5, 7 and 8 have frequencies higher than 12.3 MHz (observed in sample No.~1) for the $\pm 5/2 \leftrightarrow \pm 7/2$ transition. 
As shown in Fig.~\ref{MagneticOrderT1}, the systematic change of the temperature dependence of $1/T_1T$ is consistently reproduced by the change of the ordering temperature $\theta$ without changing the coefficient of $a$. It should be noted that $\theta$ in sample No. 1 with $T_{\rm c} \sim 4.7$ K is evaluated to be $- 1$ K, which is very close to the quantum critical point of $\theta = 0$.
When the superconductivity is suppressed by magnetic fields, $1/T_1T$ of the sample with $T_{\rm c} \sim 4.8$ K is found to continue to increase down to 1.5 K. This verifies the validity of the fitting function (not shown here). Detailed experimental results of applying magnetic fields will be summarized in a separated paper\cite{ihara2}. 

\begin{figure}
\begin{center}
\includegraphics[width=8cm]{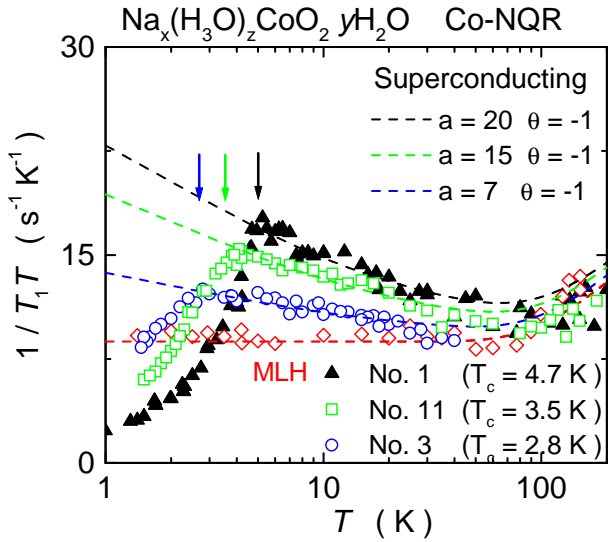}
\end{center}
\caption{Temperature dependence of $1/T_1T$ in various superconducting samples (No. 1, 3 and 11) together with that in MLH sample. Dashed lines are the fitted results. Fitting parameters for each sample are also shown (see text). }
\label{SuperconductingT1}
\end{figure}

Next, we analyze the temperature dependence of $1/T_1T$ 
in the superconducting samples using the same fitting function as above for consistency.
Figure \ref{SuperconductingT1} shows the temperature dependence of $1/T_1T$ 
in superconducting samples No.~1, No.~3 and No. 11, together with the MLH samples. 
The temperature dependences of $1/T_1T$ in these samples are satisfactorily fitted 
by the same function as described for the ordered samples using the parameters shown in the figure. 
We found that the good fit is obtained with the same $\theta=-1$ K, even if $T_{\rm c}$ of each sample is different. 
It is also noteworthy that the value of $a$ depends on $T_{\rm c}$. 
This is in contrast to the case of samples showing magnetic ordering. 
We found that the properties of the magnetic fluctuations can be systematically analyzed 
by the above fittings and that magnetic fluctuations are enhanced with increasing NQR frequencies. 
It is shown that superconductivity occurs and $T_{\rm c}$ increases 
when magnetic fluctuations with the quantum critical character appear and are enhanced at low temperatures. 
A similar conclusion is also suggested by Zheng {\it et al}. from similar analyses of $1/T_{1}T$ 
in various superconducting samples with different Na content\cite{comment}. 

We briefly comment on the character of magnetic fluctuations that is suggested by the results of NMR and NQR experiments. 
It is suggested that the $q$-dependence of magnetic fluctuations has a peak in the small $q$-vector region, 
since $1/T_1T$ at the O site is scaled to that at the Co site \cite{ning1,ihara3}. 
If there were magnetic correlations with ${\bm Q}=(\pi,\pi)$, 
the magnetic correlations at the O site would be filtered out due to the symmetry of the O position. 
In addition, from the temperature dependence of the spin part of the Knight shift, 
which shows a gradual increase with decreasing temperature but is not proportional to the temperature dependence 
of $1/T_1T$ below 30 K, 
it is considered that low-temperature fluctuations developing below 70 K have incommensurate fluctuations 
with a peak at $q \sim 0$ other than $q = 0$\cite{ihara3}.
In the next section, we discuss the relationship between the superconductivity and the magnetic fluctuations 
in the BLH compounds.

\subsection{Relationship between superconductivity and magnetic fluctuations}\label{relationship}
As described in the above section, we show that magnetic fluctuations are sample-dependent and 
seem to be related to the superconductivity. 
The clarification of the relationship between the superconductivity and magnetic fluctuations 
is very important for understanding the mechanism of the superconductivity, 
because magnetic fluctuations are considered to suppress conventional superconductivity, 
but to give rise to unconventional superconductivity. As shown in Fig. \ref{SuperconductingT1}, 
the higher-$T_{\rm c}$ sample has the stronger magnetic fluctuations. 
It should be noted that the sample with the stronger magnetic fluctuations shows a sharp decrease of $1/T_1T$ just below $T_{\rm c}$. 
It is considered that the magnitude of the superconducting gap, $2\Delta/k_{\rm B}T_{\rm c}$, is larger, 
and the superconductivity therefore is stronger, in the sample with stronger magnetic fluctuations. 
In addition, the sample quality is considered to be nearly equivalent in all BLH samples, 
on the basis of the measurement of the FWHM of the NQR spectrum. 
We consider that these experimental results support the idea that 
the magnetic fluctuations are the driving force of the superconductivity of the BLH compound. 
If magnetic fluctuations were unfavorable for superconductivity, 
the decrease of $1/T_1T$ would be moderate just below $T_{\rm c}$ and 
the magnitude of $2\Delta/k_{\rm B}T_{\rm c}$ would be smaller in the sample with the stronger magnetic fluctuations 
because of the pair breaking effect due to magnetic fluctuations. 
It is also noteworthy that magnetic fluctuations related to the superconductivity are not observed 
in the MLH compound or anhydrate Na$_{0.35}$CoO$_2$, but observed only in the BLH compounds. 
The drastic change in the electronic state is suggested to be caused by the further intercalation of water 
from the MLH structure to the BLH structure. 
We also found that magnetic fluctuations in the BLH compounds are very sensitive to samples. 
Next, we show the phase diagram in the BLH compounds using the NQR frequencies, 
and discuss the possible scenario of the superconductivity in BLH cobaltate. 
 
\subsection{Phase diagram of BLH compounds and possible scenario of superconductivity}

\begin{figure}
\begin{center}
\includegraphics[width=8cm]{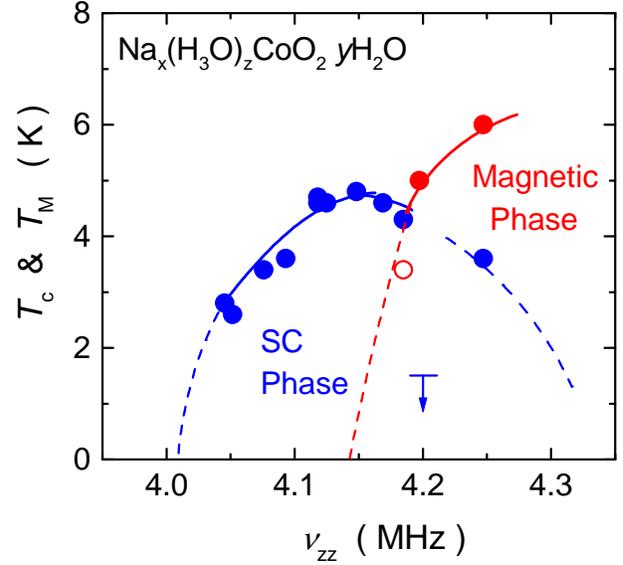}
\end{center}
\caption{Phase diagram of Na$_x$(H$_3$O)$_z$CoO$_2 \cdot y$H$_2$O, 
in which $T_{\rm c}$ and $T_{\rm M}$ are plotted against $\nu_{\rm zz}$. 
The sample with the highest $\nu_{\rm zz}$ shows a magnetic anomaly at 6 K and superconductivity at 3.5 K. 
The fitted result of $\theta$ for sample No. 7 is also shown by the red open circle. }\label{PhaseDiagram}
\end{figure}%

As shown and discussed above, 
it was found that the NQR frequency of the Co site is very sensitive to samples, 
and seems to reflect the important change of the Co electronic state. 
In the previous paper, 
we plotted $T_{\rm c}$ against the NQR frequencies arising from the $\pm 5/2 \leftrightarrow \pm 7/2$ transitions and 
developed a possible phase diagram for Na$_x$(H$_{3}$O)$_{z}$CoO$_2\cdot y$H$_2$O\cite{ihara1, iharaLT}. 

Figure \ref{PhaseDiagram} shows the new phase diagram prepared using the experimental data of twelve samples, 
where the horizontal axis is $\nu_{\rm zz}$. 
It is found that the superconducting phase shows the ``dome'' behavior against $\nu_{\rm zz}$, 
and that the highest $T_{\rm c}$ is observed at the point where $T_{\rm M}$ becomes zero, 
{\it i.e.}, at the quantum critical point.
The phase diagram we develop is reminiscent of that for Ce-based heavy-fermion superconductors, 
where antiferromagnets change to superconductors upon pressure application\cite{ce,SKawasaki}.  
We consider that the superconductivity of BLH cobaltate is induced by the quantum critical fluctuations, 
similar to the heavy-fermion superconductivity\cite{ce}.  

It is shown that the occurrence of the superconductivity is highly sensitive to $\nu_{\rm zz}$ and the $c$-axis parameter.
We discuss the physical meaning of the horizontal axis of the phase diagram for the BLH-cobaltate superconductor.
We consider that the change of $\nu_{\rm zz}$ is mainly due to the change of $V_{\text{lat}}$, 
which is related to the CoO$_{2}$ block layer thickness. 
The crystal distortion along the $c$ axis splits $t_{2g}$ levels of Co-$3d$ orbitals into 
doublet $e'_{g}$ states and singlet $a_{1g}$ state. 
Therefore, $\nu_{\rm zz}$ is regarded to be the physical parameter related to the crystal-field splitting between $a_{1g}$ and $e_{g}'$. 
According to a band calculation, these two different orbitals cross the Fermi energy and form two kinds of FSs; 
six hole pockets around the $K$ points ($e_{g}'$-FS) and a large hole Fermi surface around the $\varGamma$ point ($a_{1g}$-FS)\cite{singh}. 
The existence of the $a_{1g}$-FS was confirmed by ARPES experiments, as described in \S \ref{Intro}. 
When $e_{g}'$ levels are raised and reach the Fermi energy to form $e_{g}'$-FSs via crystal distortion and 
further crystal field splitting, 
it is considered that holelike $e'_g$-FSs appear and increase the volume. 

The above scenario explains the change of the magnetic fluctuations 
from the anhydrate Na$_{0.35}$CoO$_2$ and the MLH compound to the BLH compound.
As shown in Fig. \ref{BLHMLHT1}, 
the temperature dependences of $1/T_1T$ in the anhydrate and MLH compounds are nearly the same, 
and are governed by the $a_{1g}$-FS. 
The Korringa behavior and the pseudogap behavior are also suggested theoretically 
on the basis of the properties of the magnetic fluctuations originating from the $a_{1g}$-FS\cite{yada}. 
The low-temperature behavior of $1/T_1T$ in the BLH compounds, 
which is expressed as $a/\sqrt{T-\theta}$, is considered to be due to the $e'_g$-FSs. 
We point out that the magnetic fluctuations in the BLH compounds are well interpreted 
in terms of the multiband properties because two different temperature dependences are observed in $1/T_1T$. 
In addition, 
we found that $1/T_1T$ in the samples with lower $T_{\rm c}$ and lower $\nu_{\rm zz}$ shows a weaker temperature dependence, 
and $1/T_1T$ in the samples with higher-$T_{\rm c}$ and higher $\nu_{\rm zz}$ becomes enhanced below 70 K.
The temperature dependence of $1/T_1T$ indicates that the magnetic fluctuations at around $T_{\rm c}$ are enhanced 
with increasing $T_{\rm c}$, as shown in Fig.~\ref{SuperconductingT1}. 
We suggest that the development of magnetic fluctuations at around $T_{\rm c}$ is explained by the change in the volume of the $e'_g$-FS 
caused by the small crystal distortions along the $c$ axis. 
On the other hand, the samples with $\nu_{zz}$ higher than 4.18 MHz (No.~5, 7, and 8) show superconductivity 
at temperatures lower than 4.6 K, 
although magnetic fluctuations in these samples are stronger than those in No.~1 with $T_{\rm c} = 4.7$ K, 
as shown in Fig.~\ref{MagneticOrderT1}. 
In these samples, it is shown that magnetic ordering occurs at $T_{\rm M}$. 
(Magnetic ordering is expected in No.~7 when superconductivity is suppressed by a magnetic field.) 
When magnetic ordering occurs in these samples, 
the volume of the FS responsible for superconductivity decreases due to the opening of the magnetic gap 
associated with the magnetic ordering. 
The suppression of $T_{\rm c}$ in the higher-$\nu_{\rm zz}$ samples is interpreted in this way.   

The experimental results we show here seem to be consistently explained by the theoretical scenario 
developed by Yanase {\it et al.} and Mochizuki {\it et al.}, 
in which the magnetic fluctuations with a small characteristic wave vector $q$ are induced by the nesting of the $e_{g}'$-FSs \cite{mochizuki,yanase}.  
Although ferromagnetic fluctuations with the peak at $q = 0$ were excluded 
from the comparison between $\chi_{\rm bulk}$ and $1/T_1T$ in the superconducting compounds, 
the magnetic fluctuations with $q \sim 0$ seem to be consistent with the experimental results. 
Detailed inelastic neutron-scattering measurements are needed to reveal the character of the magnetic fluctuations in the superconducting samples.
In addition, 
a detailed ARPES experiment on superconducting Na$_{x}$(H$_{3}$O)$_{z}$CoO$_{2}\cdot y$H$_{2}$O is also desired, 
although the absence of the $e'_g$-FS has been reported on the anhydrate Na$_{0.35}$CoO$_2$. 
We suggest, on the basis of the $1/T_1T$ results shown in Fig. \ref{BLHMLHT1}, 
that the FSs on the BLH compounds are quite different from those on the anhydrate and MLH compounds.    

\subsection{Coexistence of magnetism and superconductivity}\label{coexistence}
\begin{figure}
\begin{center}
\includegraphics[width=8cm]{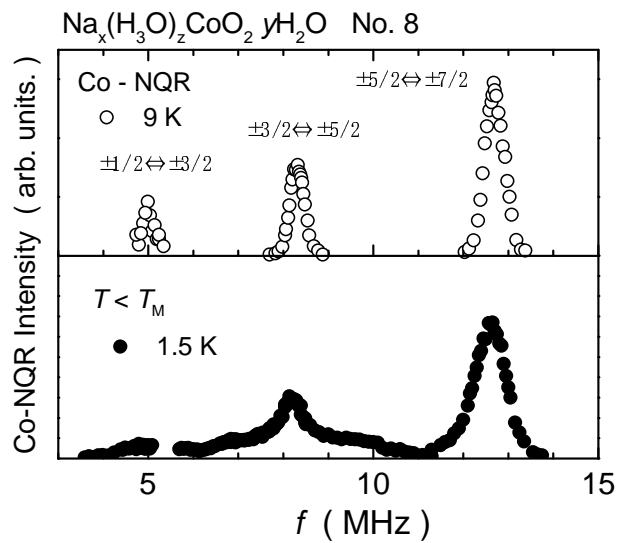}
\end{center}
\caption{
Co-NQR spectra in sample No. 8. 
Upper figure is the spectrum at 9 K above $T_{\rm M}$, and the lower figure is that at 1.5 K below $T_{\rm M}$. }
\label{SpecNo8}
\end{figure}

In samples No. 5 and 8, we observed magnetic ordering at $T_{\rm M}$. 
Detailed results for sample No. 5 have already been published in the previous paper\cite{ihara1}. 
Here, we show the phenomenon of coexisting superconductivity and magnetism observed in sample No. 8. 
As shown in Fig. 5, we found a prominent peak of $1/T_1T$ in sample No. 8 at 6 K, 
below which the internal field appears at the Co site. 
Figure \ref{SpecNo8} shows the Co-NQR spectra above and below $T_{\rm M}$. 
Three well-separated peaks above $T_{\rm M}$ become broad below $T_{\rm M}$ due to the internal field at the Co-nuclear site, 
which originates from the Co-$3d$ spins and is mediated by the hyperfine coupling between Co-$3d$ spin and Co-nuclear moment. 
Because the spectrum is structureless below $T_{\rm M}$, 
the magnitude of the internal field is considered to be distributed. 

\begin{figure}
\begin{center}
\includegraphics[width=8cm]{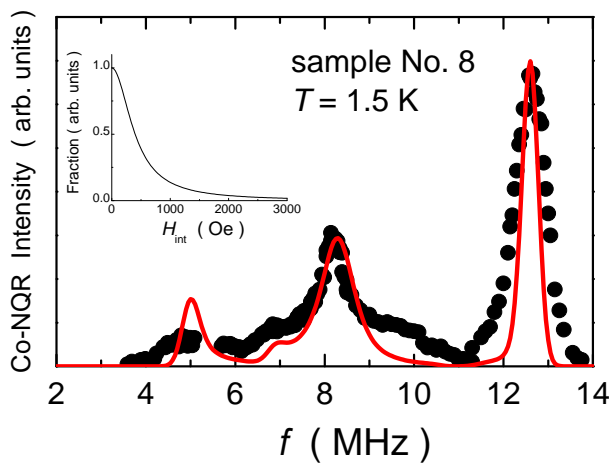}
\end{center}
\caption{
Simulation of NQR spectrum at 1.5 K. 
The red line is the calculated NQR spectrum using the inhomogeneous internal field shown in the inset.  }
\label{No8OrderSP}
\end{figure}

When the Co nuclear spins with the electric quadrupole interaction are in the magnetic field, 
the Zeeman interaction is added to the total nuclear Hamiltonian.
The Zeeman interaction is expressed as
\[
{\cal H}_{\rm Z} = -\gamma_{\rm n}\hbar I\cdot H_{\rm int},
\]
where $\gamma_{\rm n}$ and $H_{\rm int}$ are the Co nuclear gyromagnetic ratio and the internal field at the Co nuclear site, respectively. 
The structureless broad spectrum is approximately reproduced using the inhomogeneous internal field at the Co nuclear site, 
as shown in the inset of Fig. \ref{No8OrderSP}. 
Here, the magnitude of the internal field is distributed, 
suggesting inhomogeneous ordered moments lying in the CoO$_2$ plane. 

\begin{figure}
\begin{center}
\includegraphics[width=8cm]{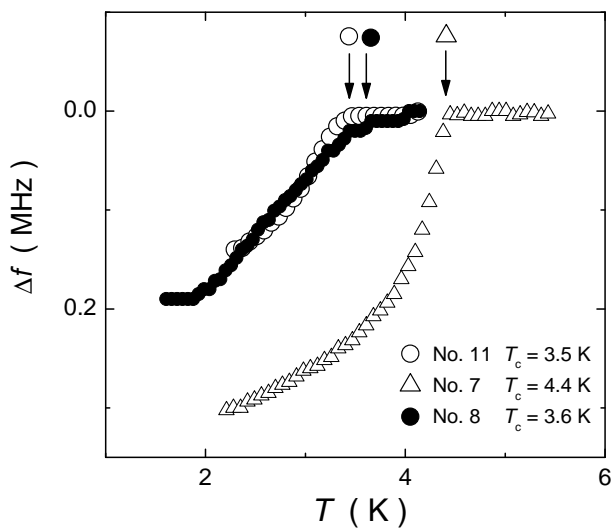}
\end{center}
\caption{
The ac-susceptibility measurements on the samples No. 7, 8, and 11 using the in-situ NQR coil. 
The superconducting transition on No. 8 is broad, but the Meissner signal of No. 8 is comparable to that on No. 11, 
which has a similar $T_{\rm c}$.  }
\label{MeissnerNo8}
\end{figure}

Below 3.5 K, Meissner shielding was observed. 
The superconducting transition of sample No. 8 is broader, 
but the magnitude of Meissner shielding is comparable to that in other samples, as shown in Fig. \ref{MeissnerNo8}.
The temperature dependence of $1/T_1T$ measured at 12.69 MHz did not show the clear anomaly associated with the superconducting transition at 3.5 K. 
The recovery of the nuclear magnetization after the saturation pulses is consistently fitted by the single component of $T_1$ at 12.69 MHz. 
We also measured $1/T_1T$ at various frequencies around 8.5 MHz, 
where the internal field is distributed from zero to 1 kOe, as shown in Fig. \ref{No8OrderSP}. 
The values of $1/T_1T$ at 1.5 K are nearly the same for these frequencies. 
These experimental results rule out the macroscopic phase separation between superconducting and magnetic regions, 
but suggest microscopic coexistence of superconductivity and magnetism.
However, taking into account that the static moments are distributed in the magnetically ordered state, 
we consider that superconductivity and magnetism coexist in the inhomogeneous magnetic structure; 
superconductivity occurs in the smaller magnetic moment region and magnetism occurs in the larger moment region. 
We point out that such coexistence state is observed in CeCu$_2$(Si$_{1-x}$Ge$_{x}$)$_2$ 
with small Ge concentration ($0<x<0.05$)\cite{YKawasaki}.
We suggest that the inhomogeneous coexistence is the characteristics of the compounds whose $T_{\rm M}$ is higher than $T_{\rm c}$.
To investigate the relationship between superconductivity and magnetism in Na$_x$(H$_3$O)$_z$CoO$_2 \cdot y$H$_2$O, 
further experiments are still needed. In particular, pressure experiments on sample No. 8 are considered to be very important. 

\section{Conclusion}

We performed Co-NQR measurements on various samples with different values of $T_{\rm c}$ and $T_{\rm M}$. 
We found that the nuclear quadrupole frequency $\nu_{\rm Q}$ strongly depends on the sample.
We show that the asymmetric parameter $\eta$ is almost unchanged ($\eta \sim$ 0.2) in the samples, 
and suggests that the presence of holes in both $a_{1g}$ and $e'_{g}$ orbitals is important in understanding the finite $\eta$, 
although the ordering of Na and oxonium ions in the Na-hydrate block layer should be taken into account. 
In contrast to the constant $\eta$, 
the electric field gradient $\nu_{\rm zz}$ depends on the samples and has a linear relation with the $c$-axis parameter.
We consider that the sample dependence of $\nu_{\rm zz}$ originates mainly from the trigonal distortion of the CoO$_{2}$ layer, 
which splits the $t_{2g}$ levels of the Co $3d$ orbital into $e_{g}'$ and $a_{1g}$ levels. 
We also show that the magnetic fluctuations, which develop below 70 K, 
are one of the characteristic features observed in the BLH compounds, 
and that these fluctuations are strongly related to the occurrence of superconductivity.   
We found that the magnetic fluctuations are also related to $\nu_{\rm zz}$, 
and suggest that $\nu_{\rm zz}$ is one of the most important physical parameters 
for characterizing the sample properties of superconducting and non-superconducting Na$_{x}$(H$_{3}$O)$_{z}$CoO$_{2}\cdot y$H$_{2}$O compounds.
We developed a phase diagram using the experimental results of our twelve samples, 
and showed that the high-$T_{\rm c}$ samples are situated near the point where $T_{\rm M}$ becomes zero. 
The microscopic coexistence of the superconductivity and magnetism is suggested for sample No. 8, 
which is similar to that in Ce-based superconductors.
We conclude that BLH cobaltate is an unconventional superconductor induced by quantum critical fluctuations, 
and that it shares many aspects with Ce-based heavy-fermion superconductors. 

\begin{acknowledgments}
We thank Y. Itoh, H. Ohta, H. Yaguchi, S. Nakatsuji and Y.~Maeno for support in the experiments and for valuable discussions. 
We also thank S.~Fujimoto, K.~Yamada, Y.~Yanase, M. Mochizuki, and M.~Ogata for valuable discussions.
This work was partially supported by CREST of the Japan Science and Technology Agency (JST) and the 21 COE program on 
``Center for Diversity and Universality in Physics'' from MEXT of Japan, and by Grants-in-Aid for Scientific Research 
from the Japan Society for the Promotion of Science (JSPS) (No. 16340111 and 18340102) and MEXT (No. 16076209).
\end{acknowledgments}


\begin{thebibliography}{99}

\bibitem{takada}
K.~Takada, H.~Sakurai, E.~Takayama-Muromachi, F.~Izumi, R.~A.~Dilanian and T.~Sasaki: Nature {\bf 422} (2003) 53.

\bibitem{jorgensen}
J.~D.~Jorgensen, M.~Avdeev, D.~G.~Hinks, J.~C.~Burley and S.~Short: Phys. Rev. B {\bf B 68} (2003) 214517.

\bibitem{foo}
M.~L.~Foo, R.~E.~Schaak, V.~L.~Miller, T.~Klimczuk, N.~S.~Rogado, Y.~Wang, G.~C.~Lau, C.~Craley, H.~W.~Zandbergen, N.~P.~Ong and R.~J.~Cave: Solid State Commun. {\bf 127} (2003) 33.

\bibitem{takada2}
K.~Takada, H.~Sakurai, E.~Takayama-Muromachi, F.~Izumi, R.~A.~Dilanian and T.~Sasaki: J. Solid. State. Chem. {\bf 177} (2004) 372.

\bibitem{bayrakci}
B.~P.~Bayrakci, I.~Mirebeau, P.~Bourges, Y.~Sidis, M.~Enderle, J.~Mesot, D.~P.~Chen, C.~T.~Lin and B.~Keimer: Phys. Rev. Lett. {\bf 94} (2005) 157205.

\bibitem{helme}
L.~M.~Helme, A.~T.~Boothroyd, R.~Coldea, D.~Prabhakaran, D.~A.~ Tennant, A.~Hiess and J.~Kulda: Phys. Rev. Lett. {\bf 94} (2005) 157206.

\bibitem{Arita}
R.~Arita: Phys. Rev. B {\bf 71} (2005) 132503.

\bibitem{ishida1}
K.~Ishida, Y.~Ihara, Y.~Maeno, C.~Michioka, M.~Kato, K.~Yoshimura, K.~Takada, T.~Sasaki, H.~Sakurai and E.~Takayama-Muromachi: J. Phys. Soc. Jpn. {\bf 72} (2003) 3041.

\bibitem{ihara1}
Y.~Ihara, K.~Ishida, C.~Michioka, M.~Kato, K.~Yoshimura, K.~Takada, T.~Sasaki, H.~Sakurai and E.~Takayama-Muromachi: J. Phys. Soc. Jpn. {\bf 74} (2005) 867.

\bibitem{lynn}
J.~W.~Lynn, Q.~Huang, C.~M.~Brown, V.~L.~Miller, M.~L.~Foo, R.~E.~Schaak, C.~Y.~Jones, E.~A.~Mackey and R.~J.~Cava: Phys. Rev. B {\bf 68} (2003) 214516.

\bibitem{sakurai1}
H.~Sakurai, K.~Takada, T.~Sasaki and E.~Takayama-Muromachi: J. Phys. Soc. Jpn. {\bf 74} (2005) 2909.

\bibitem{singh}
D.~J.~Singh: Phys. Rev. B {\bf 61} (2000) 13397.

\bibitem{qian}
D.~Qian, L.~Wray, D.~Hsieh, D.~Wu, J.~L.~Luo, N.~L.~Wang, A.~Kuprin, A.~Fedorov, R.~J.~Cava, L.~Viciu and M.~Z.~Hasan: Phys. Rev. Lett. {\bf 96} (2006) 046407.

\bibitem{yang}
H.-B.~Yang, Z.-H.~Pan, A.~K.~P.~Sekharan, T.~Sato, S.~Souma, T.~Takahashi, R.~Jin, B.~C.~Sales, D.~Mandrus, A.~V.~Fedorov, Z.~Wang and H.~Ding: Phys. Rev. Lett. {\bf 95} (2005) 146401.

\bibitem{shimojima}
T.~Shimojima, T.~Yokoya, T.~Kiss, A.~Chainani, S.~Shin, T.~Togashi, C.~Zhang, C.~Chen, S.~Watanabe, K.~Takada, T.~Sasaki, H.~Sakurai and E.~Takayama-Muromachi: J. Phys. Chem. Solids, {\bf 67} (2006) 282.

\bibitem{shimojima2}
T.~Shimojima, unpublished. 

\bibitem{mochizuki}
M.~Mochizuki, Y.~Yanase and M.~Ogata: J. Phys. Soc. Jpn. {\bf 74} (2005) 1670.

\bibitem{yanase}
Y.~Yanase, M.~Mochizuki and M.~Ogata: J. Phys. Soc. Jpn. {\bf 74} (2005) 2568.

\bibitem{yada}
K.~Yada and H.~Kontani: J. Phys. Soc. Jpn. {\bf 74} (2005) 2161.

\bibitem{ihara4}
Y.~Ihara, K.~Ishida, C.~Michioka, M.~Kato, K.~Yoshimura, K.~Takada, T.~Sasaki, H.~Sakurai and E.~Takayama-Muromachi: J. Phys. Soc. Jpn. {\bf 73} (2004) 2069.

\bibitem{ihara3}
Y.~Ihara, K.~Ishida, K.~Yoshimura, K.~Takada, T.~Sasaki, H.~Sakurai and E.~Takayama-Muromachi: J. Phys. Soc. Jpn. {\bf 74} (2005) 2177.

\bibitem{ihara5}
Y.~Ihara, K.~Ishida, H.~Takeya, C.~Michioka, M.~Kato, Y.~Itoh, K.~Yoshimura, K.~Takada, T.~Sasaki, H.~Sakurai and E.~Takayama-Muromachi: J. Phys. Soc. Jpn. {\bf 75} (2006) 013708.

\bibitem{milne}
C.~J.~Milne, D.~N.~Argyriou, A.~Chemseddine, N.~Aliouane, J.~Veira, S.~Landsgesell and D.~Alber: Phys. Rev. Lett. {\bf 93} (2004) 247007.

\bibitem{sakurai4}
H.~Sakurai, N.~Tsujii, O.~Suzuki, H.~Kitazawa, G.~Kido, K.~Takada, T.~Sasaki and E.~Takayama-Muromachi, Phys. Rev. {\bf B 74} (2006) 092502.

\bibitem{takada3}
K.~Takada, K.~Fukuda, M.~Osada, I.~Nakai, F.~Izumi, R.~A.~Dilanian, K.~Kato, M.~Takata, H.~Sakurai, E.~Takayama-Muromachi and T.~Sasaki: J. Mat. Chem. {\bf 14} (2004) 1448.

\bibitem{zheng1}
G.-q.~Zheng, K.~Matano, R.~L.~Meng, J.~Cmaidalke and C.~W.~Chu: J. Phys. Condens. Matter {\bf 18} (2006) L63.

\bibitem{sakurai2}
H.~Sakurai, K.~Takada, T.~Sasaki, F.~Izumi, R.~A.~Dilanian and E.~Takayama-Muromachi: J. Phys. Soc. Jpn. {\bf 73} (2004) 2590.

\bibitem{sakurai3}
H.~Sakurai, K.~Takada and E.~Takayama-Muromachi: to be published in {\it Progress in Superconductivity Research} (Nova Science Publishers). 

\bibitem{Hanzawa}
K.~Hanzawa, F.~Komatsu and K.~Yosida: J. Phys. Soc. Jpn. {\bf 59} (1990) 3345.  

\bibitem{zheng2}
G.-q.~Zheng, Y.~Kitaoka, K.~Ishida and K.~Asayama: J. Phys. Soc. Jpn. {\bf 64} (1995) 2524.

\bibitem{Ray}
R.~Ray, A.~Ghoshray, K.~Ghoshray and S.~Nakamura: Phys. Rev. B {\bf 59} (1999) 9454.

\bibitem{Mukhamedshin}
I.~R.~Mukhamedshin, H.~Alloul, G.~Collin and N.~Blanchard: Phys. Rev. Lett. {\bf 94} (2005) 247602.

\bibitem{ning1}
F.~L.~Ning and T.~Imai: Phys. Rev. Lett. {\bf 94} (2005) 227004.

\bibitem{kobayashi1}
Y.~Kobayashi, H.~Watanabe, M.~Yokoi, T.~Moyoshi, Y.~Mori and M.~Sato: J. Phys. Soc. Jpn. {\bf 74} (2005) 1800.

\bibitem{fujimoto}
T.~Fujimoto, G.-q.~Zheng, Y.~Kitaoka, R.~L.~Meng, J.~Cmaidalka and C.~W.~Chu: Phys. Rev. Lett. {\bf 92} (2004) 047004.

\bibitem{ning2}
F.~L.~Ning, T.~Imai, B.~W.~Statt and F.~C.~Chou: Phy. Rev. Lett. {\bf 93} (2004) 237201.

\bibitem{Yokoi}
M.~Yokoi, T.~Moyoshi, Y.~Kobayashi, M.~Soda, Y.~Yasui, M.~Sato and K.~Kakurai: J. Phys. Soc. Jpn. {\bf 74} (2005) 3046.

\bibitem{Moriya}
T.~Moriya: J. Mag. Mag. Mat. {\bf 100} (1991) 261.

\bibitem{ihara2}
Y.~Ihara, to be published.

\bibitem{comment}
Zheng {\it et al.} analyzed the temperature dependence of $1/T_1T$ using the formula 
$(1/T_1T) = c/(T+\theta) + (1/T_1T)_0$\cite{zheng1}. 
The first term on the right hand side is the expression anticipated in the case of the 2-D nearly AFM compounds 
within the SCR theory. 
The main difference between our analyses and theirs is that we determined  $\left(1/T_1T\right)_{\rm MLH}$ 
from the temperature dependence of $1/T_1T$ in the higher temperature region, 
whereas $(1/T_1T)_0$ is the fitting parameter in their analysis. 
This difference between the analyses is considered to be due to the different temperature dependence of $1/T_1T$; 
$1/T_1T$ in the higher temperature region seems to depend on the Na content.   

\bibitem{iharaLT}
Y.~Ihara, H.~Takeya, K.~Ishida, C.~Michioka, K.~Yoshimura, K.~Takada, T.~Sasaki, H.~Sakurai and E.~Takayama-Muromachi: To appear in AIP.

\bibitem{ce}
N.~D.~Mathur, F.~M.~Grosche, S.~R.~Julian, I.~R.~Walker, D.~M.~Freye, R.~K.~M.~Haselwimmer and G.~G.~Lonzarich: Nature {\bf 394} (1998) 39.

\bibitem{SKawasaki}
S.~Kawasaki, T.~Mito, G.-q.~Zheng, C.~Thessieu, Y.~Kawasaki, K.~Ishida, Y.~Kitaoka, T.~Muramatsu, T.~C.~Kobayashi, D.~Aoki, Y.~Haga, A.~Araki, R.~Settai and Y.~Onuki: Phys. Rev. B {\bf 65} (2002) 20504.

\bibitem{YKawasaki}
Y.~Kawasaki, K.~Ishida, K.~Obinata, K.~Tabuchi, K.~Kashima, Y.~Kitaoka, O.~Trovarelli, C.~Geibel and F.~Steglich: Phys. Rev. B {\bf 66} (2002) 224502.

\end{thebibliography}
\end{document}